\documentclass[12pt]{iopart}
\usepackage{iopams}
\usepackage{textcomp}
\usepackage{bbold}
\expandafter\let\csname equation*\endcsname\relax
\expandafter\let\csname endequation*\endcsname\relax
\usepackage{amsmath}
\usepackage{cite}
\usepackage{textcomp}
\usepackage[breaklinks=true,colorlinks,citecolor=blue,linkcolor=blue,urlcolor=blue]{hyperref}
\usepackage{graphicx}
\usepackage{dcolumn}
\usepackage{bm}
\usepackage{xcolor}
\usepackage[normalem]{ulem}
\newcommand{\ket}[1]{|#1\rangle}

\begin{document}
\title[Topological nano-switches]
{Topological nano-switches in higher-order topological insulators}
\author{Joseph Poata$^1$, Fabio Taddei$^2$, and Michele Governale$^1$}
\address{
$^1$ School of Chemical and Physical Sciences and MacDiarmid
Institute for Advanced Materials and Nanotechnology, Victoria University of
Wellington, PO Box 600, Wellington 6140, New Zealand}
\address{$^2$ NEST, Instituto Nanoscienze-CNR and Scuola Normale Superiore, I-56126, Pisa, Italy}

\begin{abstract}
We consider multi-terminal transport through a flake of rectangular shape of a two-dimensional topological insulator in the presence of an in-plane magnetic field.  This system has been shown to be a second-order topological insulator, thus exhibiting corner states at its boundaries. The position of the corner states and their decay length can be controlled by the direction of the magnetic field. 
In the leads we assume that
the magnetic field is absent and therefore we have helical one-dimensional propagating
states characteristic of the spin-Hall effect.
 Using a low-energy effective Hamiltonian we show analytically that, in a two-terminal setup, transport can be turned on and off by a rotation of the in-plane magnetic field.
Similarly, in a three terminal configuration, the in-plane magnetic field can be used to turn on and off the transmission between neighbouring contacts, thus realising a directional switch.
Analytical calculations are supplemented by a numerical finite-difference method.
For small values of the Fermi energy and field strength, the analytical results agree exceptionally well with the numerics. The effect of disorder is also addressed in the numerical approach.
 We find that the switching functionality is remarkably robust to the presence of strong disorder stemming from the topological nature of the states  contributing to the electron transport.
\end{abstract}

\maketitle
\submitto{\NJP}

\section{Introduction}
Recently the family of topological materials has grown to include what are known as higher-order topological insulators (TIs)\cite{Benalcazar2017,Benalcazar2017PRB,Song2017,Langbehn2017,Schindler2018,Geier2018} (see Ref.~\cite{Xie2021} for a review).
These systems are characterized by the presence of gapless boundary states in $d-n$ dimensions, where $d$ is the system dimensionality and $n$ is an integer less than or equal to $d$.
In particular,
second-order topological insulators (SOTIs) present low-energy conducting one-dimensional states at the boundaries between different surfaces (hinge states) if $d=3$, or zero-dimensional bound states at the corners (corner states) if $d=2$.
Here we focus on the latter case.
So far only a few materials have been proposed to realise a SOTI in two dimensions (2D). These are twisted bilayer graphene~\cite{Park2019}, graphdiyne~\cite{Sheng2019}, breathing Kagome lattices~\cite{Ezawa2018b}, phosforene~\cite{Ezawa2018c}, and cubic semiconductor quantum wells~\cite{Krishtopenko2021}.
Interestingly, SOTIs can be  implemented by applying an in-plane Zeeman (or exchange) field to 2D TIs, for example produced by a magnetic substrate, as it was proposed in Refs.~\cite{Ezawa2018,Ren2020,Chen2020,Wang2021,Wang2022,Wu2022,Long2023,Krishtopenko2024}.
Although a substantial body of literature exists already on SOTIs in 2D, few papers have addressed their transport properties, see Refs.~\cite{Ezawa2018,Wang2021,Wang2022,Wu2022,Long2023}.

In the present paper we concentrate on flakes of SOTI in the form of a rectangle.
For these systems it has been shown that the probability density of the corner states can be changed by rotating the flake with respect to the crystal axes (see Ref.~\cite{Poata2023} for SOTIs belonging to the Cartan class IIIA).
In particular, when the flake of SOTI is obtained from a 2D TI with an in-plane Zeeman field, any convex polygon always exhibits two corner states~\cite{Poata2023}. 
Considering a rectangle, the two corner states are localized at its vertices unless the Zeeman field is parallel to an edge.
In this case, the corner states extend along the edges parallel to the field becoming effectively ``edge states''.
More precisely, the decay length of the probability density of the corner states diverges along the edges which are parallel to the Zeeman field, while remaining finite along the edges perpendicular to it.
This effect can be exploited to control the electronic transport through a flake attached to leads by rotating the external magnetic field for a given crystal axes orientation.
This has been numerically shown in Ref.~\cite{Ezawa2018} using a toy model for squares and hexagons defined on a lattice and where the leads were simple 1D trivial chains.

In this paper we calculate the conductance of a system consisting of a rectangular flake of SOTI made of a 2D TI subjected to an in-plane Zeeman field, and attached to either two or three leads in which the field is absent, see Fig.~\ref{fig:setup}.
In particular, we study the behaviour of the scattering amplitudes as functions of the direction of the Zeeman field.
These are first calculated analytically by considering the effective low-energy Hamiltonian for the edges~\cite{Poata2023} and by matching the wavefunctions of the relevant edges of the systems.
In the two-terminal setup we find that the conductance (proportional to the probability for electrons to be transmitted from one lead to another) presents an on/off switching behaviour as a function of the field direction, peaking at a maximum value of ($2e^2/h$) when the Zeeman field is directed horizontally (the {\it on state} of the switch).
Indeed, for this field orientation the boundary state extends along the entire horizontal edge, thus connecting the two leads. 
On the contrary, when the field direction deviates from the horizontal, the corner states are strongly localized in the regions close to the leads and transport is inhibited -- the {\it off state} of the switch.
We find that the angular width of the peaks decreases linearly with the length $L$ of the flake and with the strength of the Zeeman field.
For larger values of $L$ the peaks acquire  additional features which reflect the occurrence of Fabry-P\'erot-type resonance within the flake.
Remarkably, in the leads, reflection occurs between edge states belonging to the same edge,  and the edge states localised on opposite edges of the TI remain uncoupled.
This is possible since the presence of a Zeeman field in the flake breaks time-reversal symmetry, allowing back scattering on the same edge of the 2D TI.
The two terminal device realises a topologically protected \emph{on/off switch} controlled by the direction of the magnetic field.

We find an analogous phenomenology in the three-terminal setup for the conductance between leads 1 and 3, see Fig.~\ref{fig:setup}.
A remarkably similar behaviour is also found for the conductance between leads 1 and 2, the peaks now corresponding to the field pointing in the vertical direction, as it is for this direction of the field that the boundary states extend along the entire vertical edge.
The system thus behaves like a \emph{directional switch}.

The analytical scattering amplitudes are then compared to the exact results obtained by using the full Hamiltonian discretised on a square grid. The numerical calculations were performed using the Kwant code \cite{Kwant2014}. 
Indeed, for both setups we find that the conductances are exceptionally well approximated  by the analytical model as long as
the Fermi energy is much smaller than the bulk gap and the Zeeman energy is smaller than the bulk gap.
When these two conditions are not fulfilled substantial discrepancies occur between the analytics and numerics.
Finally, we address the effect of disorder in the flake by including an on-site random potential.
As expected, disorder suppresses the interference effects, thus removing the additional oscillations produced by Fabry-P\'erot resonance.
Remarkably, however, we find that disorder does not alter the main conductance peaks even for very strong disorder, since the conducting states, which extend along an edge 
(horizontal or vertical), are not affected by scattering events which do not break time-reversal symmetry.
On the other hand, the conductance in the off state can take a finite value, but only for very high strength of disorder (of the order of the bulk gap).
We have also checked that the presence of a potential barrier at the interface between the leads and the flake has no effect on the conductance (Klein paradox).

The paper is organized as follows. In Sec.~\ref{sec:Model}, we detail the system and the model, including the effective Hamiltonian for the edges of the flake of SOTI, the scattering states in the leads and the scattering region.
The analytical and numerical results are presented and discussed in Sec.~\ref{sec:twoT}, for the two-terminal setup, and in Sec.~\ref{sec:threeT}, for the three-terminal setup.
Two additional appendices are included to describe the derivation of the effective edge Hamiltonian (\ref{app:Heff}) and the mode matching conditions to calculate the scattering amplitudes (\ref{app:modeMatching}).

\section{Model}
\label{sec:Model}
We consider the two setups shown schematically in Fig.~\ref{fig:setup}. In the first [panel a)], the scattering region is attached to two leads, whereas in the second setup [panel b)] the scattering region is attached to three leads. We assume the leads to be made of a two-dimensional TI whose low-energy physics is described by the Hamiltonian 
\begin{equation}
\label{eq:H_TI}
H_{\mathrm{TI}}= m(\hat{\mathbf{k}})\sigma_0\tau_z 
+ A (\hat{k}_x\sigma_x+\hat{k}_y\sigma_y)\tau_x ,
\end{equation}
where $\hbar\hat{\mathbf{k}}$ is the momentum operator, $\tau_i$ is the $i$-th Pauli matrix in orbital space and $\sigma_i$ the $i$-th Pauli matrix in spin-space and the operator $m(\hat{\mathbf{k}})$ is defined as 
\begin{equation}
m(\hat{\mathbf{k}})=m_0+m_2 \hat{\mathbf{k}}^2.
\end{equation}
The bulk gap in the spectrum of $H_{\mathrm{TI}}$ is 
\begin{align*}
\Delta_0=\text{min}\left[ |m_0|, \sqrt{-\frac{A^2}{m_2}\left(m_0+\frac{1}{4}\frac{A^2}{m_2}\right)} \right]. 
\end{align*}
For the sake of definiteness, we assume  $m_0<0$, $m_2>0$, and $A>0$. Since $\mathrm{sign}(m_0/A)<0$, the Hamiltonian $H_{\mathrm{TI}}$ is topologically non trivial and supports helical edge states. 

In the scattering region we assume that there is a non-vanishing magnetic field, $\mathbf{B}$, in the direction $\cos(\theta)\hat{\mathbf{x}}+\sin(\theta)\hat{\mathbf{y}}$, described by the Hamiltonian
\begin{equation}
\label{eq:H_M}
H_{\mathrm{M}}=
\Omega_Z \left(\cos(\theta)\sigma_x+\sin(\theta)\sigma_y\right)\tau_0,
\end{equation}
where $\Omega_Z>0$ is the Zeeman energy. The Hamiltonian of the scattering region is therefore $H_\mathrm{S}=H_{\mathrm{TI}}+H_{\mathrm{M}}$ and describes a SOTI~\cite{Ezawa2018,Wu2022,Long2023} in Cartan class AIII, i.e. time-reversal and charge-conjugation symmetries are both broken but there is an additional chiral symmetry. In the present case the chiral symmetry is represented by the operator $\sigma_z \tau_x$, that is $\left\{H_S, \sigma_z \tau_x\right\}=0$, with $\{\dots \}$ denoting the anticommutator. 

\subsection{Effective Hamiltonian for the edges}

\begin{figure}
    \centering
    a)\phantom{\hspace{0.55\columnwidth}}\\[6pt]
    \includegraphics[width=0.6\columnwidth]{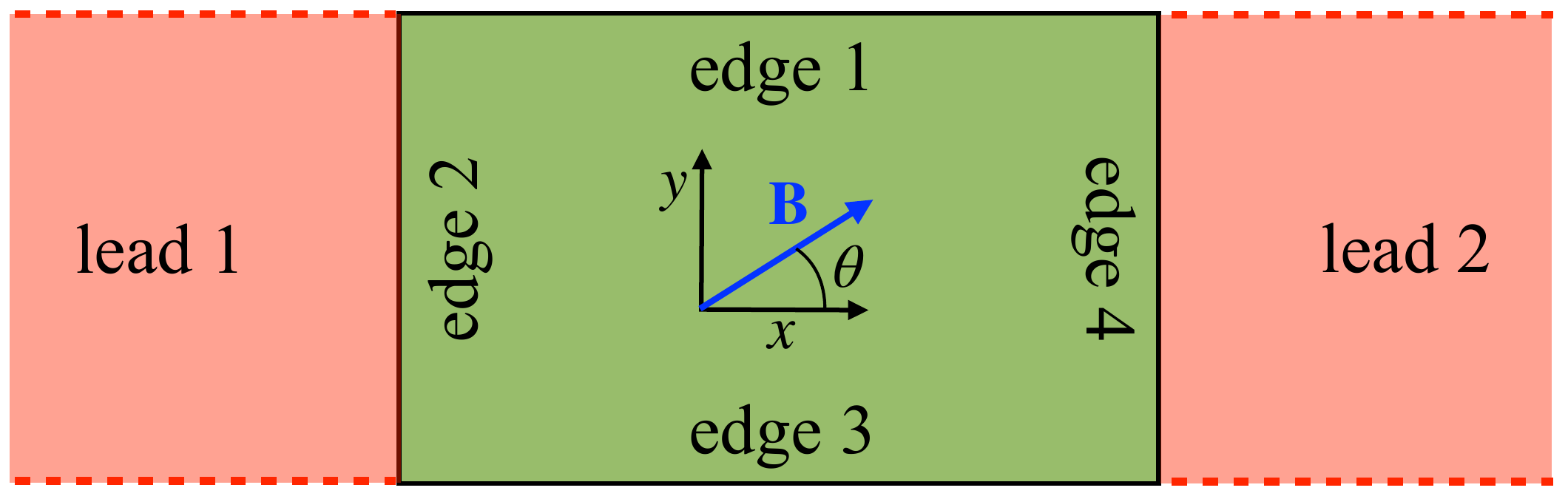}\\[3pt]
    b)\phantom{\hspace{0.55\columnwidth}}\\[-12pt]
\includegraphics[width=0.5\columnwidth]{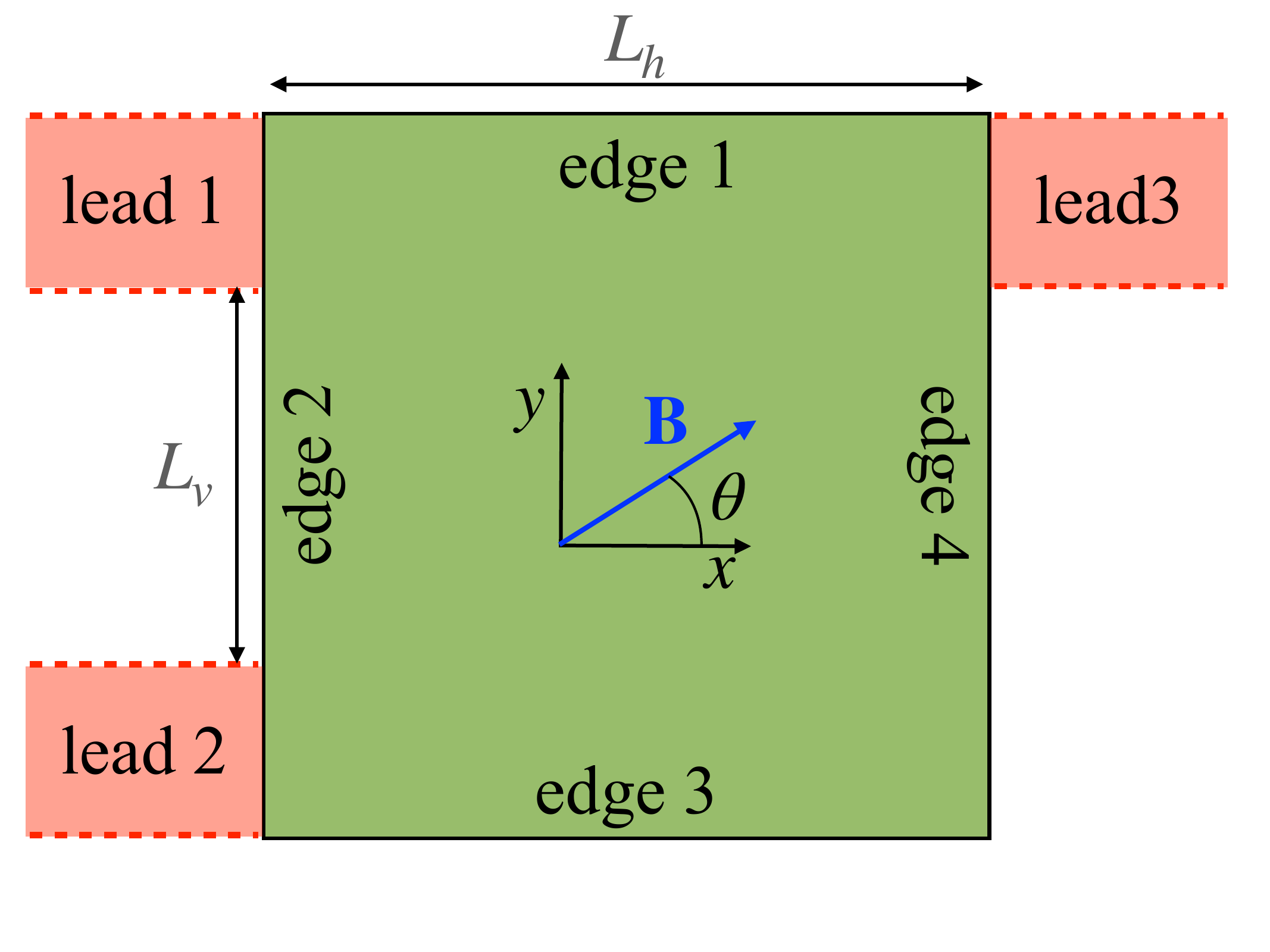}\\
    \caption{Schematic diagram of the two setups considered in this article: a) two terminal on-off switch; b) three-terminal directional switch. The scattering region (green) is made up of a TI subjected to an in-plane magnetic field $\mathbf{B}$ which gaps the helical edge states. The leads (red) are 2D TIs.}
    \label{fig:setup}
\end{figure}
The effective Hamiltonian for a linear edge in the scattering region has been derived in Ref.~\cite{Poata2023}
for $\theta=\pi/4$. For an arbitrary magnetic field direction $\theta$, the effective Hamiltonian for the edge identified by the angle $\alpha$ describes massive Dirac Fermions and it is given by (see~\ref{app:Heff})
\begin{equation}
\label{eq:H_alpha_eff}
H_{\alpha,\text{eff}}=\left(
\begin{array}{cc}
{- A \hat{k}_\parallel}   & i M(\alpha,\theta) e^{-i \alpha} \\
-i M(\alpha,\theta) e^{i\alpha}     & {A \hat{k}_\parallel}
\end{array}
\right),
\end{equation}
where the induced mass reads
\begin{equation}
\label{eq:MI}
M(\alpha,\theta)= \Omega_Z \sin(\alpha-\theta).
\end{equation}
Equation~(\ref{eq:H_alpha_eff}) is written in the basis of the helical states at $\hat{k}_\parallel=0$ for a TI ($\Omega_z=0$):  
\begin{subequations}
\label{eq:edge-states}
\begin{align}
    |\Phi_{-,\alpha}\rangle=& \frac{ \rho(x_\perp-W/2)}{\sqrt{2}}\,  \left( \ket{+,\uparrow}- e^{i\alpha} \ket{-,\downarrow}\right)\, ,\\
     |\Phi_{+,\alpha}\rangle=& \frac{\rho(x_\perp-W/2)}{\sqrt{2}}\,\left( \ket{+,\downarrow} + e^{-i\alpha} \ket{-,\uparrow} \right)\, ,
\end{align}
\end{subequations}
where we have introduced the basis $\{\ket{\tau,\sigma}\}$, with $\tau\in\{+,-\}$ and $\sigma\in\{\uparrow,\downarrow\}$. 
The function $\rho(x_\perp-W/2)$ describes the transverse profile of the edge-state wavefunction for the edge located at $x_\perp=W/2$
(see~\ref{app:Heff}). 
Here $x_{\parallel}$ and ${x}_{\perp}$ are the coordinates aligned, respectively, parallel and perpendicular (pointing outward from scattering region) to the edge.
For the setup under consideration, we have numbered the edges of the scattering region as shown in Fig.~\ref{fig:setup}. The angle $\alpha_i$ corresponding to the $i$-th edge is given by 
\begin{equation}
\alpha_i=(i-1)\frac{\pi}{2}.     
\end{equation}

\subsection{Scattering states in the leads}
Before elucidating the working principle of these devices, we need to discuss the scattering states in the leads.
These are helical in nature and localised at the two edges.
In this work, we assume that the width $W$ of the leads is much larger than the characteristic transverse decay length of the edge states, $R_0=|A/m_0|$, and therefore hybridisation between helical states on different edges can always be neglected. 
These states have a linear dispersion $E_{R/L}=\pm A k_{\text{lead}}$, where $k_{\text{lead}}$ is the momentum along $x$,
and are given by
\begin{subequations}
\begin{align}
\label{eq:scattering_states_a}
 |\Phi_{\text{R},\text{upper}}\rangle &=\frac{\rho(y-W/2)}{\sqrt{2}}\,\left( \ket{+,\downarrow} +\ket{-,\uparrow} \right)e^{i k_{\text{lead}} x},\\
 |\Phi_{\text{R},\text{lower}}\rangle &= \frac{\rho(-y-W/2)}{\sqrt{2}}\,\left( \ket{+,\uparrow} + \ket{-,\downarrow} \right)e^{i k_{\text{lead}} x},\\
 |\Phi_{\text{L},\text{upper}}\rangle &= \frac{\rho(y-W/2)}{\sqrt{2}}\,\left( \ket{+,\uparrow} - \ket{-,\downarrow} \right)e^{i k_{\text{lead}} x},\\
 \label{eq:scattering_states_d}
  |\Phi_{\text{L},\text{lower}}\rangle &= \frac{\rho(-y-W/2)}{\sqrt{2}}\,\left( \ket{+,\downarrow} -\ket{-,\uparrow} \right)e^{i k_{\text{lead}} x},
\end{align}
\end{subequations}
where the subscript R/L indicates whether the helical state is right-moving/left-moving and 
the subscript $\text{upper/lower}$ indicates whether it is localised near the upper/lower edge of the lead. The transverse coordinate in the leads is denoted by $y$ and we assume that $y=0$ is the central axis of the lead.  
The right(left)-moving states are eigenstates of the operator $\sigma_x\tau_x$ with eigenvalue $1(-1)$. Similarly, the states localised near the upper (lower) edge are eigenstates of the operator $\sigma_y\tau_y$ with eigenvalue $1~(-1)$.

\subsection{The scattering region and the switching mechanism}
\begin{figure}
    \centering
\includegraphics[width=0.52\columnwidth]{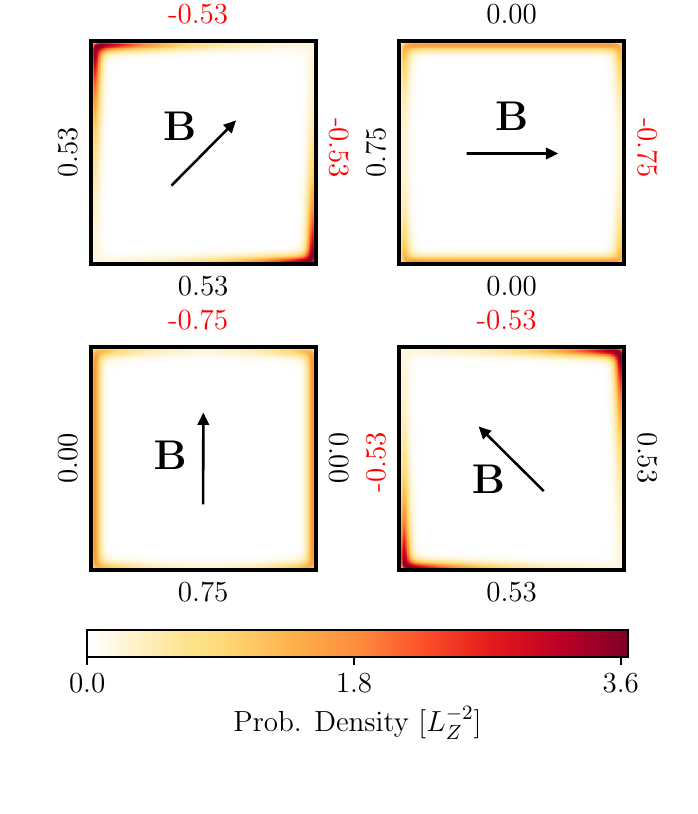}\\
    \caption{Probability density of the zero-energy boundary states for different orientations of the magnetic field. $\Omega_Z=0.1|m_0|$, $L=\frac{5}{2}L_Z$ and $m_2=\frac{11}{40}|m_0| R_0^2$ were fixed in the calculation.  The edges of the scattering regions are annotated with the values of the induced mass $M(\alpha_i,\theta)$ in units of $|m_0|$, with positive (negative) values coloured black (red). The discretization constant has been set to $0.25 R_0$. }
    \label{fig:corner-states}
\end{figure}

We now focus on the scattering region. The Hamiltonian $H_S$ describes a 2D SOTI and hence zero-energy corner states are present. In particular the corner states are located at the intersection of two edges with opposite signs of the induced mass $M(\alpha_i,\theta)$~\cite{Poata2023}.
The decay length of a corner state along edge $i$ is given by 
\begin{align*}
    R_i(\theta)=\frac{A}{|M(\alpha_i,\theta)|}.
\end{align*}
There are orientations of the magnetic field $\theta$ for which the induced mass vanishes along an edge. 
This is elucidated in Fig.~\ref{fig:corner-states}, where we see how the magnetic field direction is used to move the corner states and to generate a zero-energy boundary state that extends along an entire edge. 
As we shall justify in the following, zero-bias transport between two contacts occurs only when the extended zero-energy boundary state behave as a bridge along the edge connecting those contacts, and is a property of the device protected by its topological character.

Since it is important to assess the robustness of the devices with respect to disorder, we add to the scattering region the following Hamiltonian describing spatially uncorrelated disorder 
\begin{align}
H_{\text{disorder}}= V_0(x,y)\tau_0 \sigma_0,    
\end{align}
with $V_0(x,y)$  uniformly distributed in the range $[-V_{0, \rm{max}},V_{0, \rm{max}}]$. In the 
numerical calculations performed by discretising the model on a grid, this implies adding to each site an onsite random potential (proportional to the identity both in spin and orbital subspaces) uniformity distributed in $[-V_{0, \rm{max}},V_{0, \rm{max}}]$.

\section{Two-terminal setup}
\label{sec:twoT}
In this section we consider the two terminal setup shown in Fig. \ref{fig:setup}(a). We calculate the scattering amplitudes between different edge states (modes) in the leads as a function of the magnetic field direction $\theta$. The matrix $\mathbf{s}_{m,n}(E,\theta)$ contains all the scattering amplitudes at energy $E$ from lead $n$ to $m$, where in this case both $m$ and $n$ can only take the values 1, and 2.
The modes in the leads are listed according to their eigenvalue $\lambda$ of  $\sigma_y\tau_y$, with the first mode corresponding to $\lambda=1$ (upper edge) and the second mode  to $\lambda=-1$ (lower edge). 

The scattering problem can be solved analytically by mode matching, as detailed in \ref{app:modeMatching}. We first show the general structure of the scattering matrix.  The blocks of the scattering matrices describing transmission between different leads for the two terminal device are given by 
\begin{align}
\mathbf{s}_{2,1}(E,\theta)=\mathbf{s}_{1,2}(E,\theta)=
\left(
\begin{array}{cc}
t(E,\theta) & 0\\
 0 & t(E,\theta)
\end{array}
\right).    
\end{align}
For the reflection blocks of the scattering matrix, we find 
\begin{align}
\mathbf{s}_{1,1}(E,\theta)=-\mathbf{s}_{2,2}(E,\theta)=
\left(
\begin{array}{cc}
r(E,\theta) & 0\\
 0 & -r(E,\theta)
\end{array}
\right).    
\end{align}
The fact that the sub-blocks $\mathbf{s}_{m,n}(E,\theta)$ are diagonal tells us that the upper and lower edges are completely independent transport channels. Interestingly, when $r(E,\theta)\ne 0$ an incoming quasiparticle in the upper(lower) edge gets reflected on the same edge. This is possible because the scattering region breaks the symmetries of the TI (in particular it breaks time-reversal symmetry). 

The transmission and reflection amplitudes have compact analytical expressions: 
\begin{align}
\label{eq:2T-t}
t(E,\theta) =& \frac{k L_Z} { k L_Z\cos(k L)-i\frac{E}{\Omega_Z}\sin(k L)}\\
\label{eq:2T-r}
r(E,\theta) =&  \frac{-\sin(\theta)\sin(k L)}{k L_Z \cos(k L)-i\frac{E}{\Omega_Z}\sin(k L)},
\end{align}
where 
\begin{align*}
L_Z=\frac{A}{\Omega_Z},    
\end{align*}
and $k$ is defined as 
\begin{align}
\label{eq:k}
  k  =\frac{1}{L_Z}\sqrt{\frac{E^2}{\Omega_Z^2}-\sin(\theta)^2}
\end{align}
and is imaginary if $|E|<|\Omega_Z \sin(\theta)|$. 
The linear conductance of the system  at zero-temperature is given by the Landauer-B\"uttiker formula:
\begin{align}
\label{eq:linearG-2T}
G(E_F,\theta)=\frac{e^2}{h}\text{Tr}\left[\mathbf{s}_{2,1}(E_F,\theta)^\dagger\mathbf{s}_{2,1}(E_F,\theta)\right]=
2 G_0  \frac{E_F^2-\Omega_z^2\sin(\theta)^2}{E_F^2-\Omega_Z^2 \sin(\theta)^2\cos\left(\frac{L}{L_Z}\sqrt{\frac{E_F^2}{\Omega_Z^2}-\sin(\theta)^2}\right)^2},
\end{align}
where $G_0=e^2/h$ is the conductance quantum and $E_F$ the Fermi energy.
The conductance in  Eq.~(\ref{eq:linearG-2T}) is the central result of this section and a few comments are in order. 
\begin{enumerate}
    \item For $E_F=0$ the linear conductance simplifies to 
    \begin{align*}
       G(0,\theta)= 
       G_0 \frac{2} {\text{cosh}\left(\frac{L}{L_Z}  \sin(\theta)  \right)^2},
    \end{align*}
    which exhibits peaks for $\theta=n \pi$, with $n$ integer. The linear conductance takes the value $2G_0$ at the peaks and goes to zero away from them. The device acts therefore as an \emph{on/off switch} controlled by the direction of the magnetic field $\theta$. Whether the transmission through the device is on or off is determined by the properties of the zero-energy topological boundary states in the SOTI region. Hence, the switching mechanism is of a topological nature.   
    For $\theta$ close to $n \pi$, the linear conductance takes the form  of a Lorentzian peak 
\begin{align*}
G\approx 2 G_0\,\frac{\left(\frac{L_Z}{L}\right)^2}{(\theta-n\pi)^2+\left(\frac{L_Z}{L}\right)^2}, 
\end{align*} 
    with an angular width approximately equal to $\displaystyle\Delta\theta= 2\frac{L_Z}{L}=\frac{2 A}{\Omega_Z L}$.
    \item For $\theta=n \pi$, with $n$ integer, the linear conductance $G(E_F,n\pi)$ is equal to 2 $G_0$ for any value of $E_F$.  
    \item For $\theta=(n+1/2) \pi$, with $n$ integer, the linear conductance $G\left(E_F,(2n+1) \frac{\pi}{2}\right)\approx 0$ if  $E_F\ll \Omega_Z$ and $G\left(0,(2n+1)\frac{\pi}{2}\right)= 0$. 
\end{enumerate}

\begin{figure}
    \centering
    a)\phantom{\hspace{0.7\columnwidth}}\\[-12pt]
    \includegraphics[width=0.65\columnwidth]{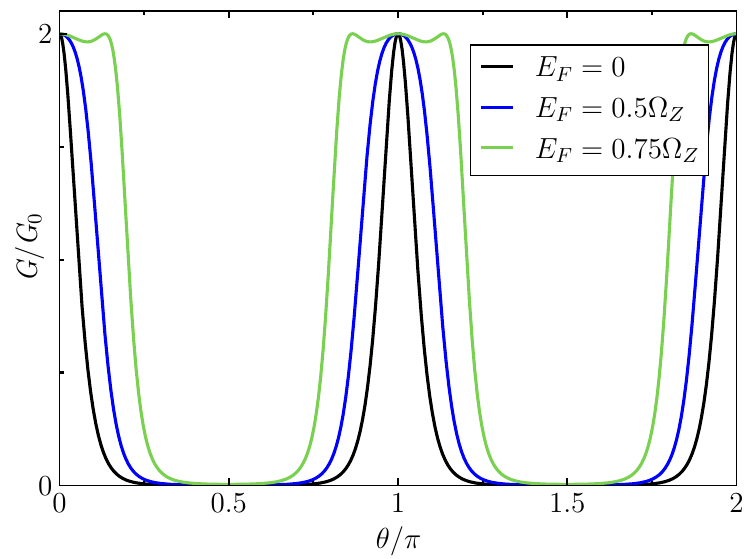}\\[3pt]
    b)\phantom{\hspace{0.7\columnwidth}}\\[-12pt]
    \includegraphics[width=0.65\columnwidth]{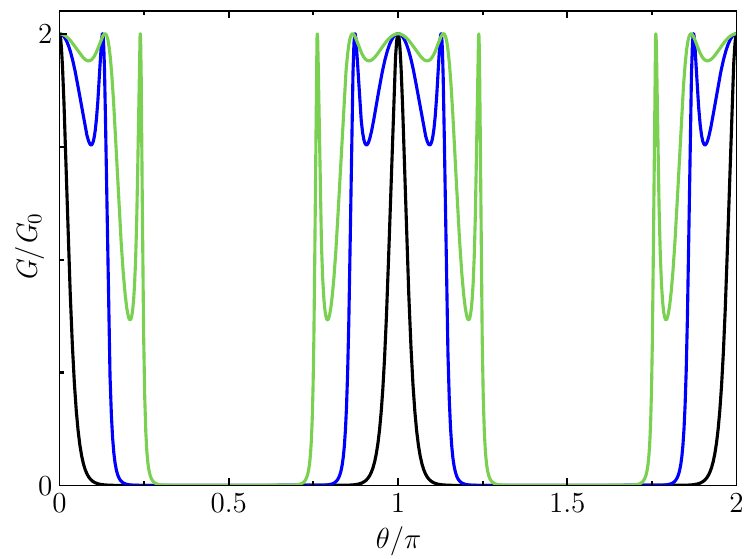}\\[3pt]
    \caption{The analytical expression for the conductance in the two-terminal setup as a function of the field direction, specified by the angle $\theta$, for three different values of the Fermi energy $E_F=0$ (black), $E_F=0.5 \Omega_Z$ (blue) and $E_F=0.75 \Omega_Z$ (green). 
    The two panels correspond to two different lengths of the scattering region: a) $L=5L_Z$ and b) $L=10 L_Z$.  }
    \label{fig:2T-conductance-q}
\end{figure}

In Fig.~\ref{fig:2T-conductance-q}, we plot the conductance in Eq.~(\ref{eq:linearG-2T}) of the two-terminal device as a function of the angle $\theta$, for different values of the Fermi energy. For $E_F=0$, that is $E_F$ in the middle of the gap, we find the peaks and switching behaviour described before. For higher value of the Fermi energy, the peaks become wider and resonances occur due to Fabry-P\'erot type of interference. This is particularly evident for the longer device $L=10\, L_Z$ of Fig.~\ref{fig:2T-conductance-q}(b). 

\begin{figure}
    \centering
    \includegraphics[width=0.65\columnwidth]{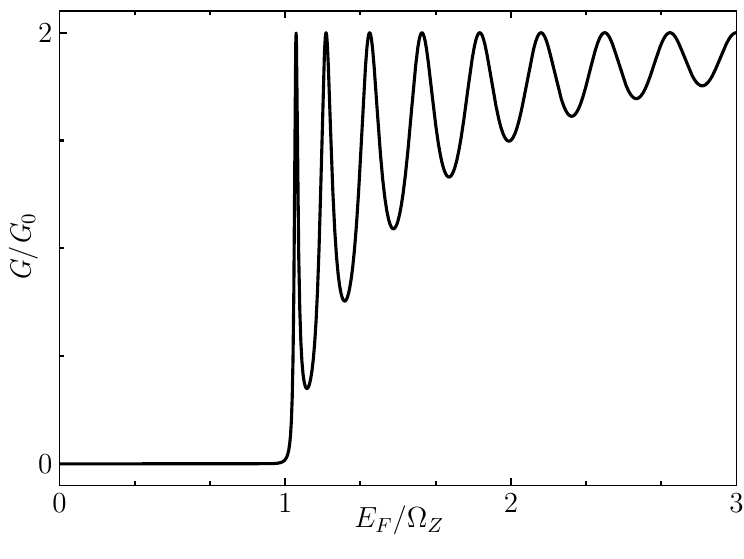}
    \caption{The analytical expression for the conductance in the off state of the two-terminal configuration, when $\theta$ equals $(n+1/2)\pi$, is displayed as a function of the Fermi energy.  The length of the scattering region is $L=10L_Z$.}
    \label{fig:2T-conductance-ef}
\end{figure}

In Fig.~\ref{fig:2T-conductance-ef}, we show the conductance of the switch in the off state, that is when $\theta=(n+1/2)\pi$. As discussed before the conductance is practically zero for $E_F<\Omega_Z$; for $E_F>\Omega_Z$ the conductance shows Fabry-P\'erot oscillations due to the finite length of the scattering region. 

\begin{figure}
    \centering
    \phantom{\hspace{0.1\columnwidth}}\\[-12pt]
    \includegraphics[width=0.65\columnwidth]{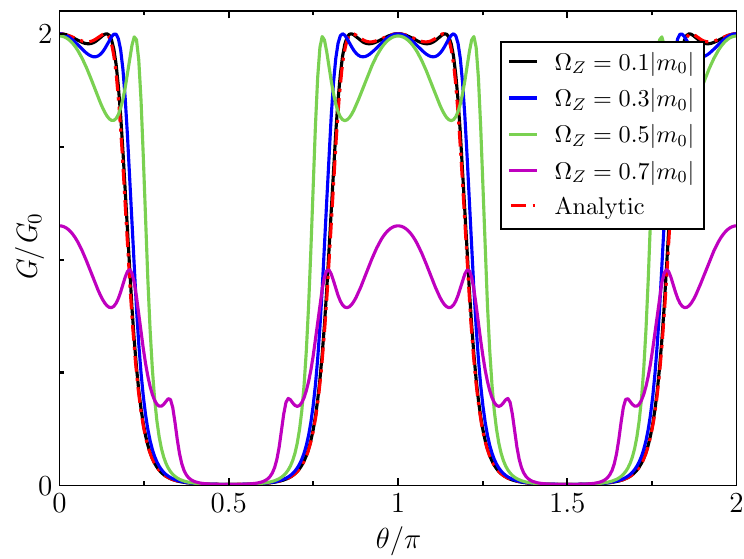}\\[3pt]
    \caption{Conductance of a two-terminal device of length $L=5L_Z$, lead width $W=5L_Z$, and fixed Fermi energy, $E_F=0.75\Omega_Z$ for different values of $\Omega_Z$. These are $\Omega_Z=0.1|m_0|$ (black), $\Omega_Z=0.3|m_0|$ (blue), $\Omega_Z=0.5|m_0|$ (green) and $\Omega_Z=0.7|m_0|$ (magenta). For the numerical calculations we have used $m_2=\frac{11}{40}|m_0|R_0^2$. The analytics are shown by the red dotted-dashed line. 
    For $\Omega_Z=0.7 |m_0|$ (magenta solid line), the condition $\Omega_Z<\Delta_0$ is violated and the peak conductance is suppressed. For the parameters used in these simulations $\Delta_0\approx 0.575 |m_0|$. Notice that since we change $\Omega_Z$ with respect to $|m_0|$ while keeping the ratio $L/L_Z$ constant, for each value of $\Omega_Z$ we take a different length $L$ of the device and a different $E_F$. This is done in order to be able to compare with the analytics, where the energy scale $|m_0|$ does not appear and the conductance depends only on the ratios $E_F/\Omega_Z$ and $L/L_Z$. The discretization constant for the numerical calculations has been set to $0.25 R_0$.}
    \label{fig:2T-conductance-omegas}
\end{figure}

The analytical model is accurate as long as $E_F\ll |m_0|$ and $\Omega_Z<\Delta_0$. Discrepancies with the results obtained using the exact numerical calculations are expected when any of these conditions is violated. This is shown in Fig.~\ref{fig:2T-conductance-omegas}, where we consider different devices with the same values of the following ratios $E_F/\Omega_Z$ and $L/L_Z$ but with different values of $\Omega_Z/|m_0|$. For all these devices, the analytical formula Eq.~(\ref{eq:linearG-2T}) predicts the same value for the conductance. However, when $E_F$ is no longer much smaller than $|m_0|$, the analytical formula starts to deviate from the numerical results. For $\Omega_Z=0.1,\ 0.3, \text{ and } 0.5\ |m_0|$, only the first condition is violated and perfect transmission at the peak is still attained. However, for $\Omega_Z=0.7 |m_0|$ (magenta solid line), the condition $\Omega_Z<\Delta_0$ is also violated (for the parameters used in the simulation $\Delta_0\approx 0.575 |m_0|$) and the peak conductance is suppressed. 

\begin{figure}
    \centering
    \phantom{\hspace{0.2\columnwidth}}\\[-12pt]
    \includegraphics[width=0.65\columnwidth]{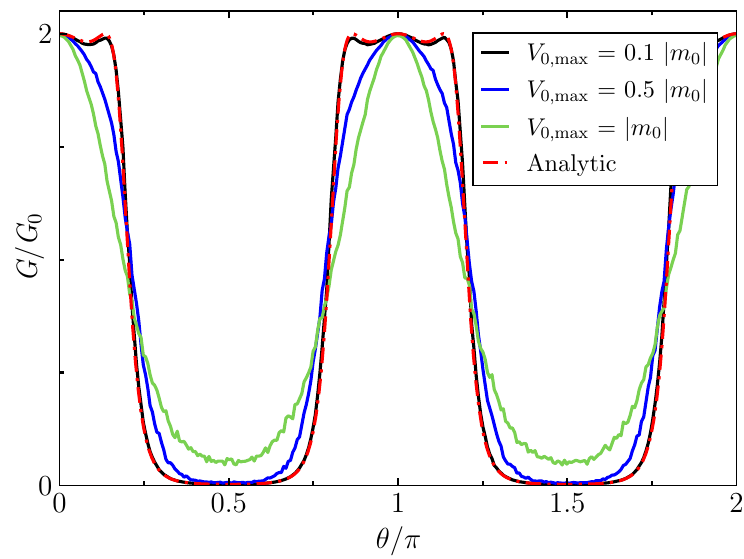}\\[3pt]
    \caption{Conductance of a two-terminal device of side length $5L_Z$, lead width
$W = 5L_Z$, and fixed Fermi energy, $E_F = 0.75\Omega_Z$, where $\Omega_Z = 0.1|m_0|$. We have also
set $m_2 = \frac{11}{40} |m_0|R^2_0$.
Different values for the strength of disorder in the scattering region are $V_{0, \rm{max}}=0.1|m_0|$ (solid black) and $V_{0, \rm{max}}=0.5|m_0|$ (blue) and $V_{0, \rm{max}}=|m_0|$ (green). The conductance has been averaged over 900 realisations of the disorder. The discretization constant for the numerical calculations has been set to $0.25 R_0$.} 
    \label{fig:2T-conductance-t-disord}
\end{figure}

We address the issue of the robustness of the topological on/off switch with respect to disorder in Fig.~\ref{fig:2T-conductance-t-disord}. 
Clearly, the disorder destroys the mesoscopic interference effects such as the Fabry-P\'erot oscillations. The perfect transmission when $\theta=n\pi$ is insensitive to the presence of disorder and survives even when $V_{0, \rm{max}}=|m_0|$. This is attributed to the fact that in the on state, transmission occurs along a TI edge state which is not affected by the disorder. However, when $V_{0,\rm{max}}\approx|m_0|$ the conductance in the off state is no longer exponentially small in $L/L_Z$, that is $\propto\exp\left[-(L/L_Z)\sqrt{1-(E_F/\Omega_Z)^2}\right]$.
This happens because of the rather large value of $E_F$ with respect to $\Omega_Z$ ($E_F=0.75 \Omega_Z$). For smaller values of $E_F$ (e.g. $\approx 0.5$ $\Omega_Z$) the conductance in the off states is much more suppressed (not shown).

Finally, it is important to notice that the presence of a potential barrier between the leads and the scattering region has no effect on the conductance due to the linear dispersion of the edge states (Klein paradox). 

\begin{figure}
    \centering
    a)\phantom{\hspace{0.7\columnwidth}}\\[-12pt]
    \includegraphics[width=0.6\columnwidth]{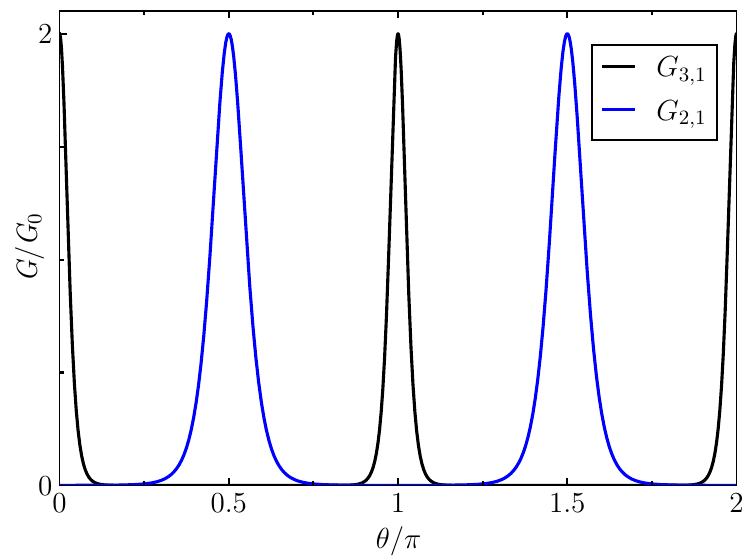}\\[3pt]
    b)\phantom{\hspace{0.7\columnwidth}}\\[-12pt]
    \includegraphics[width=0.6\columnwidth]{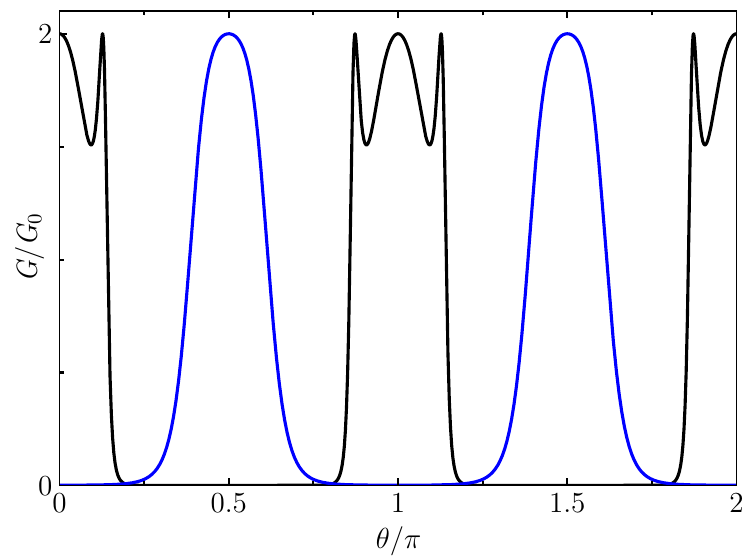}\\[3pt]
    \caption{Analytical results for the elements $G_{3,1}=G_{1,3}$ (black) and $G_{2,1}=G_{1,2}$ (blue) of the conductance matrix for the three-terminal device as a function of the angle $\theta$ for $E_F=0$ (a) and $E_F=0.5\Omega_Z$ (b). The distance between the contacts have been chosen as: $L_h=10L_Z$ and $L_v=5L_Z$.}
    \label{fig:3T-conductance-q}
\end{figure}

\section{Three-terminal setup}
\label{sec:threeT}
\begin{figure}
    \centering
    a)\phantom{\hspace{0.7\columnwidth}}\\[-12pt]
    \includegraphics[width=0.6\columnwidth]{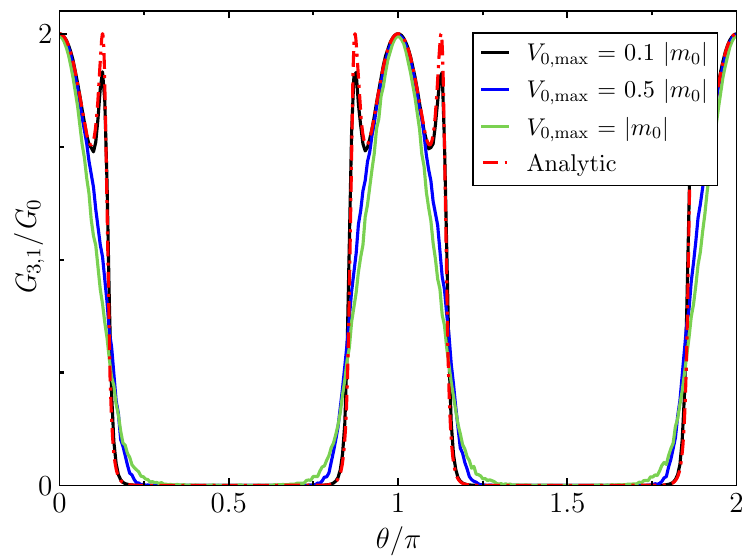}\\[3pt]
    b)\phantom{\hspace{0.7\columnwidth}}\\[-12pt]
    \includegraphics[width=0.6\columnwidth]{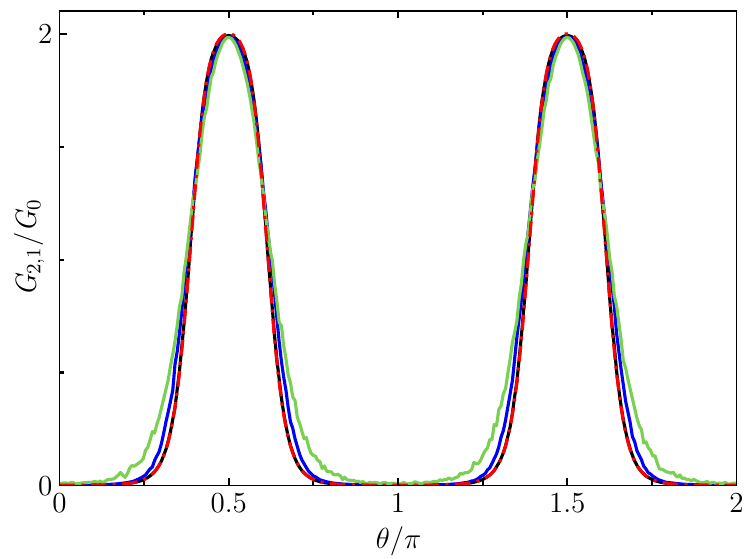}\\[3pt]
    \caption{Conductance of a three-terminal device of side lengths $L_{h}=10L_Z$ and $L_v=5L_Z$, lead width $W=2.5L_Z$, and fixed Fermi energy, $E_F=0.5 \Omega_Z$, where $\Omega_Z=0.1|m_0|$. We have also set $m_2=\frac{11}{40}|m_0|R_0^2$.
    Different values for the strength of disorder in the scattering region are $V_{0, \rm{max}}=0.1|m_0|$ (solid black) and $V_{0, \rm{max}}=0.5|m_0|$ (blue) and $V_{0, \rm{max}}=|m_0|$ (green). The conductance has been averaged over 900 realisations of the disorder. The two panels correspond to the conductance along the a) horizontal and b) vertical edges of the scattering region. The discretization constant for the numerical calculations has been set to $0.25 R_0$.}
    \label{fig:3T-conductance-t-disord}
\end{figure}

We can use the insight gained from studying transport in the two-terminal setup to construct the scattering matrix for the three terminal setup in the limit $E_F\ll|m_0|$ and $\Omega_Z<\Delta_0$. 
The induced gaps along the horizontal and vertical edges are, respectively, $M_h = -\Omega_z \textrm{sin}(\theta)$ and $M_v = \Omega_z \textrm{cos}(\theta)$.  The distance between the contacts (length travelled in the scattering region) in the $x(y)$ direction is denoted by $L_{h(v)}$ [see Fig.\ref{fig:setup} (b)].
In a similar fashion as for the two-terminal setup,  we define the reflection and transmission amplitudes along the horizontal and vertical edges as
\begin{align}
t_{h(v)}(E,\theta) =& \frac{k_{h(v)} L_Z} {k_{h(v)} L_Z\cos(k_{h(v)} L_{h(v)})-i\frac{E}{\Omega_Z}\sin(k_{h(v)} L_{h(v)})}\\
r_{h(v)}(E,\theta) =&  \frac{\frac{M_{h(v)}}{\Omega_Z}\sin(k_{h(v)} L_{h(v)})}{k_{h(v)} L_Z\cos(k_{h(v)} L_{h(v)})-i\frac{E}{\Omega_Z}\sin(k_{h(v)} L_{h(v)})} ,
\end{align}
where 
\begin{align*}
k_{h(v)} =\frac{1}{L_Z}\sqrt{\frac{E^2}{\Omega_Z^2}-\frac{M_{h(v)}^2}{\Omega_Z^2}}.
\end{align*}

Transport between lead 1 and 3 occurs along the upper edge where the scattering problem is the same as the for the two-terminal device. The corresponding block of the scattering matrix reads
\begin{align}
\mathbf{s}_{3,1}(E,\theta)=\mathbf{s}_{1,3}(E,\theta)=
\left(
\begin{array}{cc}
t_{h}(E,\theta) & 0\\
 0 & 0
\end{array}
\right).    
\end{align}
The transmission between lead 1 and 2 occurs only between the lower-edge modes of lead 1 and the upper-edge modes of lead 2. We also assume that the spinor follows adiabatically when the helical state changes direction between horizontal and vertical propagation. This allows us to write
\begin{align}
    \mathbf{s}_{2,1}(E,\theta)&=\mathbf{s}_{1,2}(E,\theta)^{T}=
\left(
\begin{array}{cc}
0 & t_v(E,\theta)\\
0 & 0
\end{array}
\right). 
\end{align}
The blocks of the scattering matrix describing the reflection amplitudes can be obtained by requiring the full scattering matrix to be unitary and they read
\begin{align}
\mathbf{s}_{1,1}(E,\theta)&=
\left(
\begin{array}{cc}
r_h(E,\theta) & 0 \\
0 & r_v(E,\theta)
\end{array}
\right)\, ,\\
\mathbf{s}_{2,2}(E,\theta)&=
\left(
\begin{array}{cc}
 -r_v(E,\theta)& 0 \\
 0& 1
\end{array}
\right)\, ,\\
\mathbf{s}_{3,3}(E,\theta)&=
\left(
\begin{array}{cc}
 -r_h(E,\theta) & 0\\
0 & 1
\end{array}
\right).
\end{align}
The elements of the conductance matrix for the three-terminal device can be obtained easily and read
\begin{align}
    G_{3,1}(E_F,\theta)&=G_{1,3}(E_F,\theta)=G_0  \frac{E_F^2-\Omega_z^2\sin(\theta)^2}{E_F^2-\Omega_Z^2 \sin(\theta)^2\cos\left(\frac{L_h}{L_Z}\sqrt{\frac{E_F^2}{\Omega_Z^2}-\sin(\theta)^2}\right)^2},\\
    G_{2,1}(E_F,\theta)&=G_{1,2}(E_F,\theta)=G_0  \frac{E_F^2-\Omega_z^2\cos(\theta)^2}{E_F^2-\Omega_Z^2 \cos(\theta)^2\cos\left(\frac{L_v}{L_Z}\sqrt{\frac{E_F^2}{\Omega_Z^2}-\cos(\theta)^2}\right)^2},\\
    G_{2,3}(E_F,\theta)&=G_{3,2}(E_F,\theta)=0.
\end{align}
The system works as a \emph{directional switch}: depending on the direction of the magnetic field, transport is enabled either between leads 1 and 3 or between leads 1 and 2. This is elucidated in Fig.~\ref{fig:3T-conductance-q}(a) where the linear conductances $G_{3,1}$ and $G_{2,1}$ are plotted as a function of the angle $\theta$ for $E_F=0$. 
The conductance $G_{3,1}$ is maximal for $\theta=n \pi$, while $G_{2,1}$ is maximal for 
$\theta=(2n+1/2) \pi$, with $n$ being an integer. The peaks of the two curves have different widths as the distances between the contacts $L_h$ and $L_v$ are in general different. The width of the peaks is given by $\Delta\theta_{h(v)}=2 L_Z/ L_{h(v)}$. The case of a finite Fermi energy is shown in Fig.~\ref{fig:3T-conductance-q}(b): as for the two-terminal device, the peaks become wider and Fabry-P\'erot resonances start to appear.
Similarly to the two-terminal case, the analytical results and the numerical results are practically indistinguishable as long as $E_F\ll |m_0|$ and $\Omega_Z<|\Delta_0|$. For larger values of the Fermi energy deviations analogous to the two-terminal case occur (not shown). 

The directional switch is also robust against disorder. This is elucidated in Fig.~\ref{fig:3T-conductance-t-disord}, where the different components of the conductance matrix are plotted as a function of the angle $\theta$ for different values of the disorder strength. Similarly to the two-terminal case, the presence of the disorder suppresses the Fabry-P\'erot oscillations, while the switching behaviour is preserved even for the greatest strength of the disorder. The fact that the behaviour of the conductance along the horizontal edge, $G_{3,1}=G_{1,3}$, is very similar to the two-terminal case is not surprising as the transport mechanism is the same. However, the situation could have been different for the conductance along the vertical edge, $G_{2,1}=G_{1,2}$. In fact, in this case we have assumed that the spinor follows adiabatically when the helical state changes direction between horizontal and vertical propagation and it was not clear a priori that this mechanism is immune to disorder. 
In contrast to Fig.~\ref{fig:2T-conductance-t-disord}, the conductance in the off-state remains small for all values of disorder considered. This is due to the fact that in Fig.~\ref{fig:3T-conductance-t-disord}, we have considered a smaller value of the Fermi energy ($E_F=0.5 \Omega_Z$).

\section{Conclusions}
In this paper we have addressed the transport through a rectangular flake of SOTI implemented by a 2D TI exposed to an in-plane magnetic field.
We have exploited the fact that the extension of the corner states (characteristic of the SOTI) along the edges of the flake is controlled by the direction of an in-plane magnetic field. In particular, when the field is aligned parallel to an edge, the corresponding corner state extends along the edge.
We have considered two setups characterized by a different number of leads where the magnetic field is absent, thus consisting of 2D TIs.
The first setup has two leads placed on opposite edges of the rectangular flake and the second one has two leads located on one edge and a third one placed on the opposite edge.
Using an effective low-energy Hamiltonian, we have calculated analytically the scattering amplitudes between the leads as a function of the direction of the magnetic field.
For both setups we have found that the zero-temperature conductance vanishes when the orientation of the field is perpendicular to an edge connecting two contacts and it exhibits  maxima for field orientations parallel to this edge.
Interestingly, the two-terminal setup realises an {\it on/off switch}, while the three-terminal setup realises a {\it directional switch}.
In addition, for large Fermi energies (and long enough edges) the main peaks are accompanied by secondary peaks which are produced by Fabry-P\'erot resonances occurring along the edges.

We have then checked these analytical results with a numerical exact approach based on the discretisation of a $4\times 4$ Hamiltonian, finding an exceptionally good agreement as long as the Fermi energy is much smaller than the bulk gap and the Zeeman energy is smaller than the bulk gap.
When these conditions are not fulfilled the conductance exhibits additional features, including a lowering of the main peaks.
Finally, the resilience of these results against the presence of disorder in the flake was addressed.
We have found that the main peaks in the zero-temperature conductance are remarkably robust to strong disorder (of the order of the bulk gap), while the resonant features are washed away. 

Fabrication of topological materials in 2D systems and the control of their topological phases are rapidly advancing, while transport setups have already been realised.
We believe that our results on transport are particularly valuable for two main reasons.
First, at a fundamental level, transport can be used as a detection tool of corner states in SOTI.
Second, as far as applications are concerned, our results can be used for designing topologically protected switches.

\section{Acknowledgements}
F.T. acknowledges financial support from the MUR - Italian Minister of University and Research - under the “Research projects of relevant national interest - PRIN 2022” - Project No. 2022B9P8LN, title “Non- equilibrium coherent thermal effects in quantum systems (NEThEQS)”, and from the Royal Society through the International Exchanges Scheme between the UK and Italy (Grant No. IEC R2 192166).

\begin{appendix}
\section{Effective edge Hamiltonian}
\label{app:Heff}
\begin{figure}[ht]
    \centering
\includegraphics[width=0.5\columnwidth]{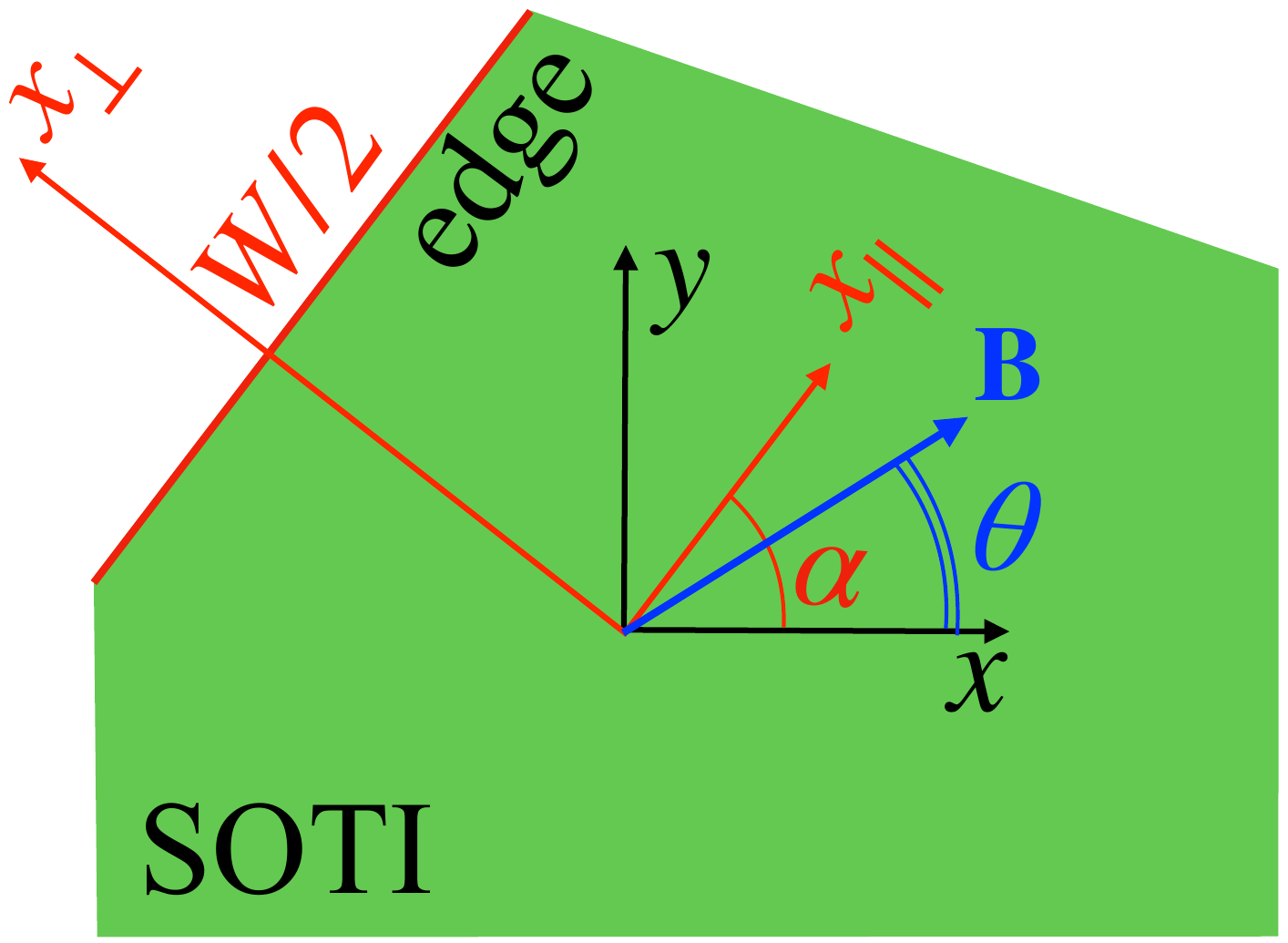}\\
    \caption{Schematic description of a linear edge (red) and of the rotated coordinate system $(x_\parallel,x_\perp)$. The edge is located at $x_\perp=W/2$, as shown. The rotation angle $\alpha$ defines the direction of the edge.}
    \label{fig:edge}
\end{figure}

In this appendix we generalise the derivation given in Ref.~\cite{Poata2023} of the effective Hamiltonian for a linear edge. We consider a linear edge at a distance $W/2$ from the coordinate origin. The direction of the edge with respect to the crystal is defined by the angle $\alpha$ by which we need to rotate the coordinate axes, so that the new coordinates, $x_\parallel$ and  $x_\perp$ are, respectively, parallel to the edge and perpendicular to it pointing outwards
(a schematic description of the edge and of the rotated coordinate system is shown in Fig.~\ref{fig:edge}). 
We take $\alpha$ in the interval $[0,2\pi)$. 
The coordinate transformation is described by
\begin{align}
\label{eq:rotation}
\left(
\begin{array}{c}
x\\
y
\end{array}
\right) 
=
\left(
\begin{array}{cc}
\cos(\alpha) & -\sin(\alpha)\\
\sin(\alpha) & \cos(\alpha)
\end{array}
\right)
  \left(
\begin{array}{c}
x_\parallel\\
x_\perp
\end{array}
\right).  
\end{align}
In the new coordinate system, the Hamiltonian (\ref{eq:H_TI}) reads
\begin{subequations}
\label{HamRot}
\begin{align}
\nonumber
H_{\text{TI}} =    m(\hat{k}_\parallel, \hat{k}_\perp)\sigma_0\tau_z +A\left[\hat{k}_\parallel (\cos(\alpha)\sigma_x+\sin(\alpha)\sigma_y)
  + \hat{k}_\perp (-\sin(\alpha)\sigma_x+\cos(\alpha)\sigma_y)\right]\tau_x ,
\end{align}
\end{subequations}
while the Hamiltonian $H_{\text{M}}$ in Eq.~(\ref{eq:H_M}) remains unchanged since this transformation (\ref{eq:rotation}) affects only the spatial coordinates and leaves the spinor basis unmodified.
We consider the semi-plane $x_\perp\in (-\infty,W/2]$, i.e. with boundary at $x_\perp=W/2$, and we start by taking $H_{\text{M}}=0$ and $\hat{k}_\parallel=0$. In this case the edge states are at zero energy and read:
\begin{align}
    \label{eq:TI-mode-1}
    |\Phi_{-,\alpha}\rangle=& \frac{ \rho(x_\perp-W/2)}{\sqrt{2}}\,  \left( \ket{+,\uparrow}- e^{i\alpha} \ket{-,\downarrow}\right)\, ,\\
    \label{eq:TI-mode-2}
     |\Phi_{+,\alpha}\rangle=& \frac{\rho(x_\perp-W/2)}{\sqrt{2}}\,\left( \ket{+,\downarrow} + e^{-i\alpha} \ket{-,\uparrow} \right)\, ,
\end{align}
where we have introduced the basis $\{\ket{\tau,\sigma}\}$, with $\tau\in\{+,-\}$ and $\sigma\in\{\uparrow,\downarrow\}$.
The function $\rho$ defining the transverse profile of the edge states is 
\begin{align}
\label{eq:rho}
\rho(x)=\frac{1}{N}\left( e^{\lambda_+ x}-e^{\lambda_- x} \right)    
\end{align}
with $N$ being a normalisation factor and 
\begin{align}
\lambda_{\pm}=\frac{A\pm\sqrt{A^2+4 m_0 m_2} }{2 m_2}. 
\end{align}
For the sake of simplicity, we assume that the parameters are such that $\lambda_{+}=\lambda_{-}^{*}$, 
with $\text{Re}(\lambda_{\pm})>0$.
Now, we include first-order linear terms in $\hat{k}_\parallel$ and $H_{\text{M}}$ as a perturbation. 
Computing the matrix elements of the perturbation on the basis $\left\{\ket{\Phi_{-,\alpha}},  \ket{\Phi_{+,\alpha}}\right\}$ we obtain the effective Hamiltonian for the edge states
\begin{align}
H_{\alpha,\text{eff}}=\left(
\begin{array}{cc}
- A \hat{k}_\parallel     & i M(\alpha,\theta) e^{-i \alpha} \\
-i M(\alpha,\theta) e^{i\alpha}     & A \hat{k}_\parallel 
\end{array}
\right),
\end{align}
where the mass term is given by
\begin{align}
M(\alpha,\theta)= \Omega_Z \sin(\alpha-\theta).
\end{align}
Such a Hamiltonian describes massive Dirac fermions with a mass term that depends on the direction of the edge, $\alpha$, and on the direction of the magnetic filed, $\theta$. 

\section{Mode matching for the two terminal device}
\label{app:modeMatching}
In this Appendix with outline the approach to calculate the scattering matrix for the two terminal device. 
If the width $W$ of the leads is much larger than $R_0$, the one-dimensional edge channels do not hibridise and can be analysed independently. For the sake of definiteness, we will consider an incoming mode on the upper edge of the left lead (lead 1). The other cases can be obtained following a very similar procedure. Making use of the expressions for the scattering states given in Eqs.~(\ref{eq:scattering_states_a}), the wavefunction in lead 1 with energy $E$, can be written as
\begin{align}
|\Psi_1(x)\rangle=\ket{\Phi_{+,0}}\,e^{i k_{\text{lead}} x}
+ r(E,\theta)\ket{\Phi_{-,0}}\, e^{-i k_{\text{lead}} x},
\end{align}
with the wavevector given by $k_{\text{lead}}=E/A$ and $\ket{\Phi_{\pm,0}}$ by Eqs.~(\ref{eq:edge-states}) with $\alpha=0$ and $x_\perp=y$. For convenience, we provide here the expressions of the states $\ket{\Phi_{\pm,0}}$: 
\begin{align*}
    \ket{\Phi_{-,0}}=& \frac{ \rho(y-W/2)}{\sqrt{2}}\,  \left( \ket{+,\uparrow}-  \ket{-,\downarrow}\right)\, ,\\
    \label{eq:TI-mode-2}
     \ket{\Phi_{+,0}}=&\frac{ \rho(y-W/2)}{\sqrt{2}} \,\left( \ket{+,\downarrow} +  \ket{-,\uparrow} \right)\, .
\end{align*}

Similarly, the wavefunction of the outgoing state in lead 2 can be written as
\begin{align}
|\Psi_2(x)\rangle=t(E,\theta)\ket{\Phi_{+,0}}\, e^{i k_{\text{lead}} (x-L)}.
\end{align}
The states in the scattering region are the eigenstates of the effective Hamiltonian Eq.~(\ref{eq:H_alpha_eff}) with $\alpha=0$ and $\hat{k}_\parallel=k$.  The wavefunction in this region at the energy $E$ can be written as 
\begin{align}
\nonumber
    |\Psi_{\text{scat}}(x)\rangle=&a_1 e^{i k x} \left[i \sin\left(\frac{\beta_k}{2}\right) \ket{\Phi_{-,0}}+  \cos\left(\frac{\beta_k}{2}\right) \ket{\Phi_{+,0}}\right]+\\ & a_2 e^{-i k x} \left[ \cos\left(\frac{\beta_k}{2}\right) \ket{\Phi_{-,0}}- i   \sin\left(\frac{\beta_k}{2}\right) \ket{\Phi_{+,0}}  \right],
\end{align}
where $k=\frac{1}{A}\sqrt{E^2-M(\theta)^2}$ with $M(\theta)=M(0,\theta)=-\Omega_Z\sin(\theta)$ and we have defined the mixing coefficients as
\begin{subequations}  
\begin{align}
    \cos\left(\frac{\beta_k}{2}\right)=&\sqrt{\frac{1}{2}\left(1+\frac{A k}{\sqrt{(A k)^2+M(\theta)^2}}\right)}\\
    \sin\left(\frac{\beta_k}{2}\right)=&\text{sign}(M(\theta))\sqrt{\frac{1}{2}\left(1-\frac{A k}{\sqrt{(A k)^2+M(\theta)^2}} \right)}.
\end{align}
\end{subequations}
Requiring the wave function to be continuous at $x=0$ and $x=L$, 
we obtain the following linear system of equations
\begin{align*}
&\cos\left(\frac{\beta_k}{2}\right)\, a_1 - i   \sin\left(\frac{\beta_k}{2}\right)\, a_2= 1\\
&i \sin\left(\frac{\beta_k}{2}\right)\, a_1 +  \cos\left(\frac{\beta_k}{2}\right) \,a_2-r(E,\theta)=0\\
&\cos\left(\frac{\beta_k}{2}\right)e^{i k L}\, a_1 - i   \sin\left(\frac{\beta_k}{2}\right)e^{-i k L}\, a_2-t(E,\theta)=0\\
&i \sin\left(\frac{\beta_k}{2}\right)e^{i k L}\, a_1 +  \cos\left(\frac{\beta_k}{2}\right)e^{-i k L} \,a_2=0.
\end{align*}
This system of equations can be easily solved and we obtain the following expressions for the reflection and transmission coefficients: 
\begin{align}
    r(E,\theta)=& \frac{M(\theta)\sin(k L)}{A k \cos(k L) - i\sqrt{A^2 k^2 + M(\theta)^2} \sin(k L )}= \frac{-\sin(\theta)\sin(k L)}{k L_Z \cos(k L)-i\frac{E}{\Omega_Z}\sin(k L)}\, ,\\
    t(E,\theta)=& \frac{A k}{A k \cos(k L) - i\sqrt{A^2 k^2 + M(\theta)^2} \sin(k L )}= \frac{k L_Z}{k L_Z \cos(k L)-i\frac{E}{\Omega_Z}\sin(k L)}\, ,
\end{align}
where in the last equality we have used $E=\sqrt{A^2 k^2 + M(\theta)^2}$ and $L_Z=A/\Omega_Z$.
\end{appendix}

\section*{References}

\bibliography{hoti}

\end{document}